\documentstyle[12pt,psfig]{article}

\def\reference{\parskip 0pt\par\noindent\hangindent 0.5 truecm}

\def\gapprox{\lower.4ex\hbox{$\;\buildrel >\over{\scriptstyle\sim}\;$}}
\def\lapprox{\lower.4ex\hbox{$\;\buildrel <\over{\scriptstyle\sim}\;$}}

\begin{document}

\title{Resonant inverse Compton scattering above polar caps:
Gap acceleration efficiency for young 
pulsars
\footnote{ADP-AT-97-4,  Publications of 
of the Astronomical Society of Australia, in press}
}

\author{Qinghuan Luo and R. J. Protheroe
} 

\date{}
\maketitle

{\center
Department of Physics and Mathematical Physics,\\
The University of Adelaide, SA 5005\\[3mm]
}

\begin{abstract}
It is shown that for moderately hot polar caps (with effective
temperature of $\sim 10^6$ K), the efficiency of polar gap acceleration
is lower compared to the case in which the polar caps are relatively
cool and inverse Compton scattering plays no role in controlling the gap.
For young pulsars with superstrong magnetic fields ($\geq10^9\ \rm T$) 
and hot polar caps (with temperature of $\ge 5\times10^6\,\rm K$), 
because of the energy
loss of electrons or positrons due to resonant inverse Compton scattering
in the vicinity of polar caps, pair cascades occur at distances further 
away from the polar cap, and in this case, we have a relatively high 
acceleration efficiency with ions carrying most of the particle luminosity.
\end{abstract}

\bigskip

\section{Introduction}

The mechanism for converting rotational energy of pulsars to
electromagnetic radiation has long been the subject of active research
in pulsar physics.  The recent observation of seven pulsars by the
instruments aboard CGRO provides even more challenge to the current
pulsar theory. These pulsars radiate high-energy photons with apparently
high efficiency (Ulmer 1994; Thompson et al. 1994). One possible
way to convert the rotational energy to
electromagnetic radiation is through particle acceleration by
the rotation-induced electric field in the pulsar's magnetosphere.
There are several sites for gap acceleration, which include the
polar gap (Ruderman \& Sutherland 1975; Arons \& Scharlemann 1979;
Michel 1974; Fawley, Arons \& Scharlemann 1977), the slot gap
(Arons 1983), and the outer gap (Cheng, Ho \& Ruderman 1986).
In the polar gap model, the acceleration region is located
near the polar cap while in the outer gap model, the acceleration
occurs near the light cylinder. In the slot gap model, the acceleration
is at the boundary between the open and closed field lines.

There is observational evidence that polar caps of young pulsars
can be hot with effective temperature $T\sim 10^6\,\rm K$
(\"Ogelman 1991; Greiveldinger et al. 1996).
Although the exact value of $T$ is rather uncertain, and may depend
on the model of the atmosphere above the polar cap (Romani 1987),
the hot polar cap appears to be a plausible consequence of polar cap heating
as the result of particle acceleration (Cheng \& Ruderman 1977; Arons
\& Scharlemann 1979; Luo 1996). One of the important
consequences of hot polar caps is that it provides
an alternative mechanism for controlling the gap, i.e.
mainly through pair cascades initiated by Compton
scattered photons (Sturner 1995; Luo 1996). Inverse Compton scattering
in the polar cap region is strongly modified by
the magnetic field in that the scattering cross section is enhanced
by the cyclotron resonance (Herold 1979), i.e. in the electron rest
frame, soft photons have energies close to the cyclotron energy
$\varepsilon_B=B/B_c$ in $m_ec^2$ with $B_c\approx4.4\times10^9\ \rm T$
the critical field. This process is referred to as resonant inverse
Compton scattering (RICS). Because of the resonance,
the effective cross section is greatly enhanced in comparison
with the ordinary Compton scattering. For gap acceleration, we
can define its efficiency as the ratio of the total voltage
across the gap to the maximum voltage across the polar cap
(i.e. for an empty magnetosphere). The efficiency
of the gap acceleration is controlled by the pair production
and thus strongly depends on the effective temperature of
the polar caps.

In this paper, we consider the constraint imposed by RICS on the
efficiency of polar gap acceleration. We assume that
free emission of charges from polar caps is allowed, and
consider both the polar region in which the outflowing charges
are electrons (${\bf B}\cdot\mbox{\boldmath{$\Omega$}}>0$) and that 
in which the outflowing charges are heavy ions or positrons
(${\bf B}\cdot\mbox{\boldmath{$\Omega$}}<0$). In the polar gap at
which ${\bf B}\cdot\mbox{\boldmath{$\Omega$}}<0$, since the energy loss 
of ions due to RICS is negligibly small,
they can be accelerated by the full potential drop across the gap.
We derive the condition for the ions to carry most
of the particle luminosity. Accelerated ions
may produce pairs in the anisotropic thermal photon field.
We calculate numerically the distance from the polar
cap at which the ions start to produce pairs and compare
it to the gap length constrained by pair production by RICS.

\section{Limits on particle energies by resonant inverse
Compton scattering}
For a hot polar cap which emits
soft photons, RICS can constrain the maximum energies of electrons
(or positrons) accelerated in the polar gap in the following two ways:
the energy loss competes with acceleration, or by limiting the acceleration
length through pair production. For a given accelerating potential, $\phi(x)$,
the Lorentz factor of an electron (or positron) at a distance, $xR_0$, to
the polar cap is given by (Dermer 1990; Luo 1996)
$$\gamma(x)=\gamma_0+{e\over m_ec^2}\,\phi(x)-
{R_0\sigma_T\over\lambda^3_c}\,{\Theta\over\pi^2
\alpha_f\beta}\int^x_0\!dx'\,\biggl[
{\xi(x')\,\varepsilon^2_B\over\gamma(x')}\biggr],\eqno(1)$$
where $\gamma_0$ is the initial Lorentz factor,
$\lambda_c=\hbar/m_ec$ is the Compton wavelength, $\Theta=1.686\times10^{-4}
(T/10^6{\rm K})$ is the effective temperature (in $m_ec^2$),
$R_0$ is the star's radius, $\sigma_T$ is the Thomson cross section,
$\alpha_f\approx1/137$ is the fine structure constant, $\xi(x')=
\varepsilon_B/[\gamma(x')(1-\beta(x')\cos\theta_{\rm max})]$, and 
$\theta_{\rm max}$ is the maximum angle between a photon and the particle
motion. For thermal emission from the entire neutron star's surface 
we have $\theta_{\rm max}=\arccos\{[(x+1)^2-1]^{1/2}/(x+1)\}$, 
which corresponds to propagation of photons at directions 
tangential to the surface. For a hot polar cap, this is
valid only for $x<0.5\theta^2_c$ (where $\theta_c=(2\pi R_0/Pc)^{1/2}$ 
and $P$ is the pulsar period); for $x>0.5\theta^2_c$, the angle
is given by $\theta_{\rm max}\approx\arccos[x/(\theta^2_c+x^2)^{1/2}]$,
corresponding to photons from the polar cap rim on which the field
lines have colatitude angle $\theta_c$ relative to the magnetic
pole. The energy loss due to RICS is described by the third term on
the right hand side of equation (1).

The maximum Lorentz factor corresponds to the value at $x=x_{\rm max}$
which is the gap length determined by the pair production, e.g.
due to RICS or due to thermal photons interacting with heavy ions,
whichever comes first. Equation (1) can
be solved numerically with the boundary condition $\gamma(0)=\gamma_0$.

The model for the accelerating potential strongly depends on
the physical conditions on the polar cap. If polar caps are
sufficiently hot, we may have free emission of charges.
The induced electric field causes charge flow,
which tends to screen out the field. However, the charge screening
is always insufficient, e.g. due to field line curvature (Arons
\& Scharlemann 1979), and hence charge depletion occurs (the charge
density deviates from Goldreich-Julian (G-J) density). An electric
potential exists in the charge-depletion region (often called the gap).
As an example, we consider the potential (e.g. Arons \& Scharlemann
1979; Arons 1983)
$$\phi(x)\approx 0.375\theta_c^4cBR_0\,\biggl({\theta\over\theta_c}
\biggr)\biggl[1-\biggl({\theta\over\theta_c}\biggr)^2
\biggr]\,g(x)\,\sin\psi\,\sin i,\eqno(2)$$
where $\theta$ is the angle of the field lines
relative to the magnetic pole, $\psi$ is the azimuthal angle
of the open field line flux tube, $x$ is the distance to the polar
cap (in $R_0$), $i$ is the angle between the rotation axis and the
magnetic pole, and $g(x)$ is given by $g(x)=x^2\ \ {\rm for}\ \
x<0.5\theta_c$, $g(x)=\theta_c\bigl[
(x+1)^{1/2}-1\bigr]\ \ {\rm for}\ \ x\gapprox0.5\theta_c$.
According to Arons (1983), the angle $\psi$ is in the ranges,
$-\pi\lapprox \psi\lapprox0$ for $i>\pi/2$ (if the outflowing charges
are ions), and $0\lapprox \psi\lapprox\pi$ for $i<\pi/2$ (if the
outflowing charges are electrons). The potential given by (2)
is due to charge density decrease along field lines (i.e. it becomes
less than the G-J density) because of field line curvature.
Plots of the Lorentz factor as a function of distance from the polar
cap are shown in Figure 1. We assume $\gamma=10$ at
$xR_0=1\ \rm cm$. As shown in the figure, energy loss due to RICS
can compete with acceleration only for high polar cap temperature
$T\gapprox5\times10^6\,\rm K$. For $T\approx 10^6\,\rm K$, the particle 
energy is constrained mainly through pair production by RICS.
\begin{figure}
\begin{center}
\psfig{file=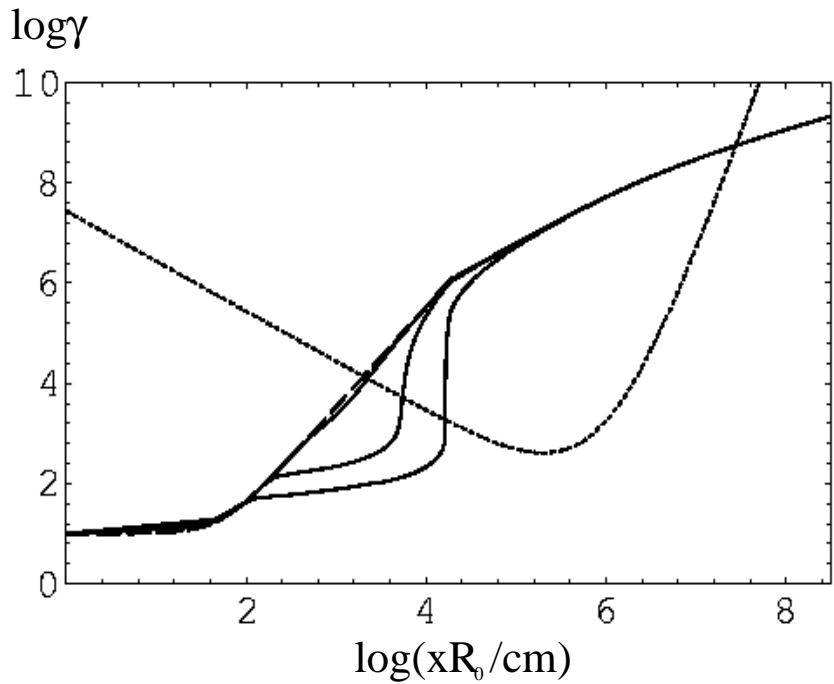,height=9cm}
\caption{ Lorentz factor versus distance to the polar cap. The electrons
(or positrons) are accelerated by the potential (2) with $i=\pi/4$ (or
$3\pi/4$), $B=3\times10^9\,\rm T$, $P=0.15\,\rm s$. The dashed curve is
obtained without RICS. The solid curves (from top to bottom) correspond to
$T=1.5\times10^6 \,\rm K$, $5\times10^6\,\rm K$, and $10^7\,\rm K$.
We assume that thermal emission is from polar cap. The dotted curve is
the threshold condition for pair production through RICS (i.e. pair
production can occur only above the curve).}
\label{figlabel}            
\end{center}
\end{figure}

There are other models of the space-charge-limited potential such as
the model discussed by Michel (1974), also by Fawley, Arons \& 
Scharlemann (1977) and Cheng \& Ruderman (1977), which do not rely on 
field line curvature and in general gives a lower voltage.
The potential discussed by Michel (1974) arises from the deviation
of the charge density from the G-J value because of particle inertia,
i.e. the velocity $v(x)$ and the magnetic field $B(x)$ have different
scaling with $x$ such that the charge density cannot be maintained at the G-J
density all the way along open field lines. The corresponding potential
can be described by (Fawley, Arons \& Scharlemann 1977)
$$\phi(x)=2x\,cB\,R_0\biggl({\Omega\over\Omega_B}\biggr)^{1/2},\eqno(3)$$
with $\Omega=2\pi/P$, $\Omega_B=ZeB/m$ and $x\lapprox\theta_c$.
For $x>\theta_c$, the electric field drops off faster than $(xR_0)^{-2}$.
The plot of the Lorentz factor as a function of $xR_0$ is shown
in Figure 2 for $T=1.5\times10^6\, \rm K$ and $T=5\times 10^6\,\rm K$.
The electric potential for $x>\theta_c$ is assumed to be constant, 
$\phi(x)=\phi(\theta_c)$. For $T=5\times 10^6\,\rm K$, the energy loss
is so severe that it continues beyond the distance $x>\theta_c$, 
resulting in a sudden decrease in $\gamma$ at $x=\theta_c$. 
As in Figure 1, for moderately hot polar caps 
($T\approx10^6\,\rm K$), the particle energy is constrained by 
electron-positron cascades started by RICS.
\begin{figure}
\begin{center}
\psfig{file=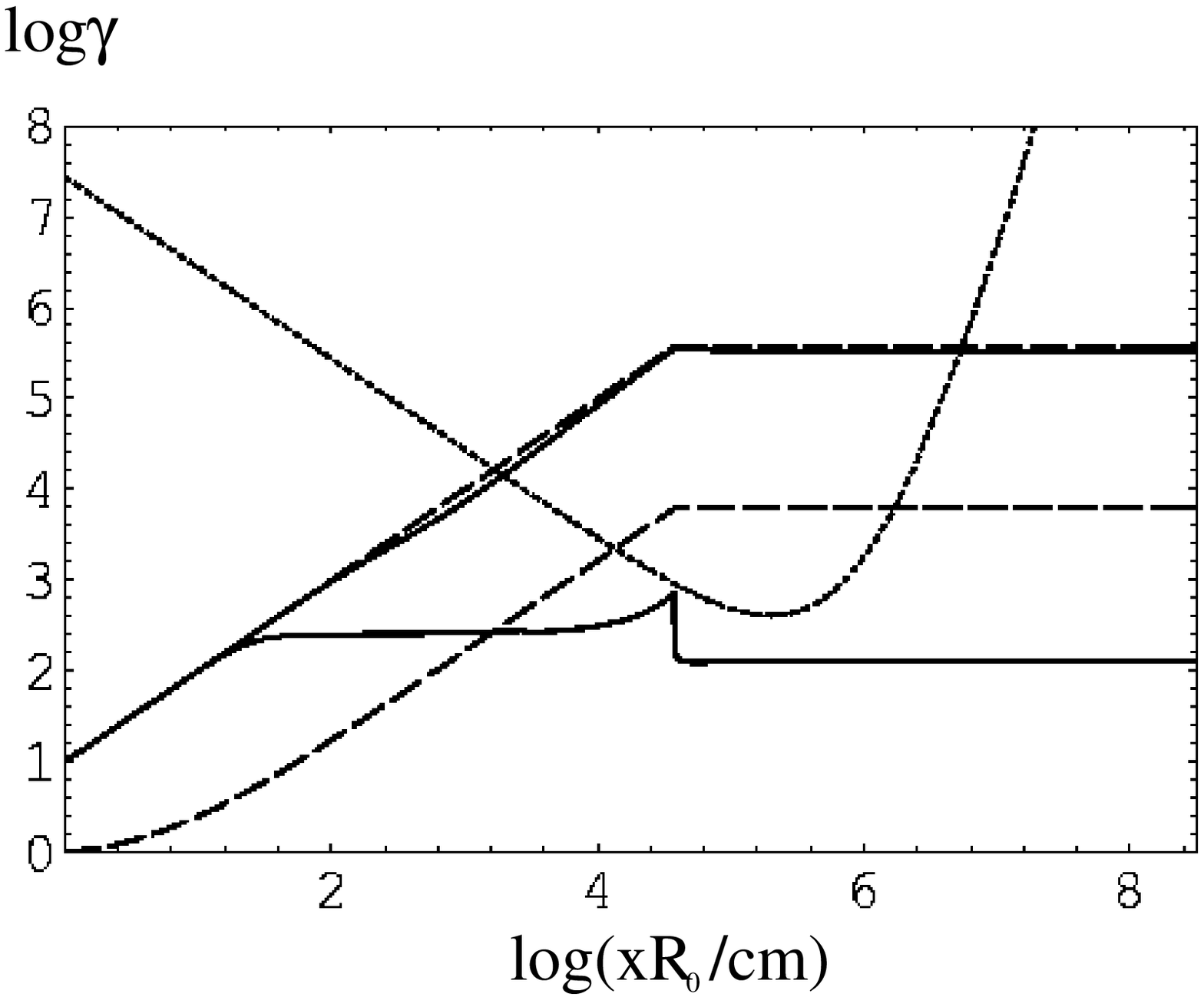,height=9cm}
\caption{As Figure 1 but the particles are accelerated by the potential
(3) with $i=0$ ($i=\pi$ for outflowing positrons or ions),
$P=0.15\,\rm s$, $B=3\times10^9\,\rm T$,
$T=1.5\times10^6\,\rm K$ (the upper solid curve), $5\times10^6\,\rm K$
(the lower solid curve). The result is similar to that derived
by Sturner (1995). The two dashed plots (upper and lower)
correspond respectively to acceleration of electron (positron)
and ion without RICS. The dotted curve is the threshold condition for
pair production through RICS.}
\label{figlabel}
\end{center}
\end{figure}

The main mechanism for a high energy photon (emitted by an electron
or positron either through curvature or inverse Compton scattering)
to produce a pair is  single photon decay in a strong magnetic field.
To produce a pair, the photon must satisfy the threshold condition
$0.5\varepsilon\sin\theta_\gamma\gapprox1$ where
$\varepsilon$ is the photon energy (in $m_ec^2$), $\theta_\gamma$ is
the angle between the photon propagation and the magnetic field.
When this condition is satisfied, the number of pairs produced
depends on the opacity, $\tau$, which is an integration of
the absorption coefficient
(for a photon being converted into a pair) along
the distance that photon travels. We assume that a photon produces
at least one pair if $\tau\gapprox1$. With this condition,
we can write down the threshold condition for producing at least
one pair through RICS (Luo 1996) as
$$\gamma>13.4\biggl({15\over\ln\Lambda}\biggr)
\biggl({B\over 3\times10^9\,{\rm T}}\biggr)^{-2}
\biggl({P\over 0.1\,{\rm s}}\biggr)^{1/2}{(1+x)^6\over x}.\eqno(4)$$
The right-hand side of (4) strongly depends on $B$.
If the above condition is satisfied, pair cascade can be initiated by 
RICS. The threshold condition is represented by the dotted curves in 
Figures 1 and 2.  Note that for each curve of $\log\gamma$ vs.
$\log(xR_0/{\rm cm})$, the location of an intersect (with the dotted 
curves) point defines a gap length, represented by $x^{_{\rm RICS}}_0$.
(If the curve has two intersecting points, the one that is the nearest
to the polar cap is relevant.) 

\section{Pair production by photons in the field of the nucleus}

Electron positron pair cascades can be started by accelerated
ions in thermal photon fields. A photon can decay into an electron 
positron pair in the Coloumb field of a nucleus when the center of 
momentum frame energy exceeds the rest mass of the nucleus plus two 
electrons, that is
$$ s > (m_N c^2 + 2 m_e c^2)^2,\eqno(5)$$
where $m_N$ is the mass of the nucleus.
This condition can be satisfied for soft photons from the polar cap 
in the presence of accelerated ions.  We can estimate the mean free 
path of an ion before producing an average  one pair. If we believe
one pair per primary particle is enough to short out the electric 
field, this mean free path length, represented by $x^{\rm ion}_0$,
can be identified as the gap length (Figures 3 and 4).
\begin{figure}
\begin{center}
\psfig{file=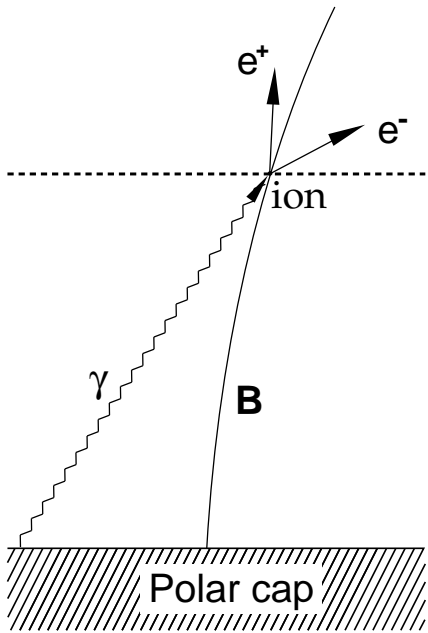,height=9cm}
\caption{Pair production due to ion colliding with soft photon
from polar cap emission above the polar cap with
$\mbox{\boldmath{$\Omega$}}\cdot{\bf B} <0$. Beside ions, the outflowing
charges may include positrons.}
\label{figlabel}
\end{center}
\end{figure}

\begin{figure}
\begin{center}
\psfig{file=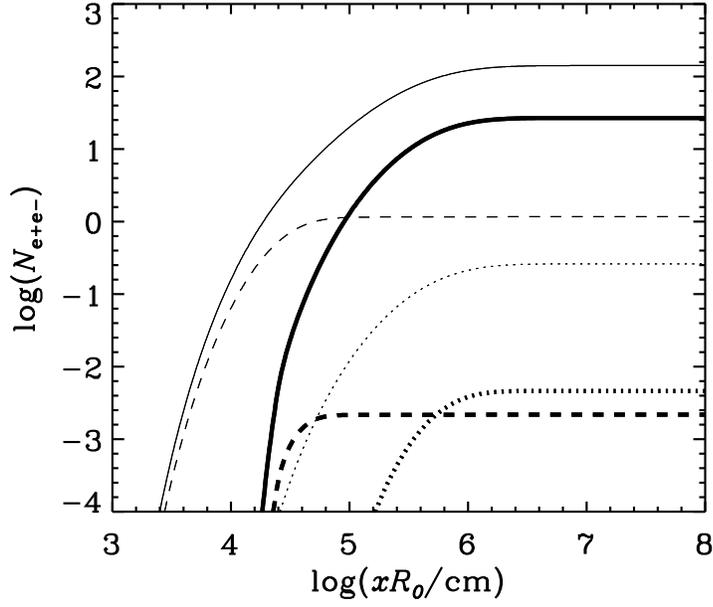,height=9cm}
\caption{Plots of cumulative number of pairs produced per ion
as a function of distance. Solid, dashed and dotted curves
correspond respectively to $T=5\times10^6\,\rm K$ for the whole star's
surface, $T=10^7\,\rm K$ for the polar cap (but the star's surface
has $T=10^6\,\rm K$), and $T=5\times10^6\,\rm K$ for the polar cap (but
the star's surface has $T=5\times10^5\,\rm K$). The thick and thin
curves are calculated using the potentials (2) and (3), respectively.
We assume $P=0.15\,\rm s$, $B=3\times10^9\,\rm T$.
}
\label{figlabel}
\end{center}
\end{figure}

Numerical evaluation of the mean free path for an accelerated ion 
traversing an isotropic thermal photon field was given by Protheroe (1984), 
and Protheroe \& Johnson (1996). The injection rate of pairs for a given 
energy of iron nuclei was numerically calculated by Bednarek \& Karakula 
(1995) for an anisotropic photon field from the polar cap.  We calculate 
number of pairs produced by iron nuclei accelerated along open field lines 
by the potentials given by Eqs. (2) and (3), and the mean free path corresponds 
to the average distance at which one pair is produced per ion nucleus.
The mean cumulative number of pairs produced in the radiation field of 
the polar cap (or whole neutron star) at temperature $T$ by a nucleus of 
energy $E$ traveling along the neutron star magnetic axis from the pole 
to distance $xR_0$ above the polar cap surface is given by,
$$N_{e^+e^-}(E) \approx 
	{R_0 \over 2} \int_0^x 
	\int_0^{\infty} n(\varepsilon)
   \int_{\cos[\theta_{\rm max}(x)]}^{\cos[\theta_{\rm min}(E,\varepsilon)]}
	(1 - \beta \cos \theta) 
        \sigma(s) \, d\cos\!\theta \, d\varepsilon \, dx,
 \eqno(6)$$
where $n(\varepsilon)$ is the differential photon number density
of photons of energy $\varepsilon$ in black body radiation at
temperature $T$, and $\sigma(s)$ is the total cross section for
pair production (Protheroe 1997) by a the nucleus at center of
momentum (CM) frame energy squared, $s$, given by
$$s=m_N^2 c^4 + 2 \varepsilon E(1 - \beta \cos \theta)
\eqno(7)$$
where $\theta$ is the angle between the directions of the nucleus 
and photon.  Note that $\theta_{\rm min}(E,\varepsilon)$ is determined 
by the threshold condition (5). Plots of accumulative number of
pairs produced per ion through interaction with thermal photons
as a function of distance from the polar cap are shown in Figure 4.
for $P=0.15\,\rm s$ and $B=3\times10^9\,\rm T$. Compared to
RICS, in this case we have $x^{_{\rm RICS}}_0<x^{\rm ion}_0$.
Although we use a high $B$, the process strongly depends on $P$ and $T$. 

\section{Acceleration efficiency}

The efficiency of the polar gap is described by the
ratio of the potential drop across the gap, $\phi(x_0)$,
to the maximum potential across the polar cap with an empty
magnetosphere, $\phi_{\rm max}$, viz.
$$\eta={\phi(x_0)\over\phi_{\rm max}},\eqno(8)$$
where $\phi_{\rm max}\!=0.5\theta^4_{\!c}cBR_0\!\approx\!6.6\times10^{12}\,
{\rm V}\, (1\,{\rm s}/P)^2(B/10^8\,{\rm T})$,
$\theta_c\!=\!(\Omega R_0/c)^{1/2}=
1.45\times10^{-2}\, (1\,{\rm s}/P)^{1/2}$, and $x_0$ is the length
of the polar gap in units of $R_0$.  Let $L_{\rm p}$ and $L_{_{\rm R}}$
be the accelerated particle and spin-down luminosities, respectively.
If the energy loss of accelerated particles in the gap
is not important, we have $\eta\approx L_{\rm p}/L_{_{\rm R}}$ since
we can write $L_{\rm p}\approx Ze\phi(x_0)\dot{N}_{_{\rm GJ}}$ and
$L_{_{\rm R}}\approx Ze\phi_{\rm max}\dot{N}_{_{\rm GJ}}$
where $\dot{N}_{_{\rm GJ}}$ is the injection rate calculated from the G-J
density $n_{_{\rm GJ}}\approx2\epsilon_0B\Omega/(Ze)$.
Note that the efficiency defined by
(8) is similar to that used by Arons (1996). For the presently
available models for accelerating potentials, we always have $\eta<1$.

Using (8), we may estimate the efficiency for a given model
potential. For the potential described by Eq. (2) and assuming
$x_0\gg0.5\theta_c$, we have the maximum efficiency
$$\eta_{\rm max}\approx0.29\theta_cx^{1/2}_0\sin\psi\,\sin i.
\eqno(9)$$
Thus, even when we use $x_0\sim R_L/R_0=1/\theta^2_c$, the maximum efficiency
is $\eta_{\rm max}\lapprox0.29$.  When a pair cascade occurs, we
usually have $x_0\ll 1/\theta^2_c$ (for the polar gap) and
hence $\eta\ll\eta_{\rm max}$.

One may estimate the efficiency for the potential given by Eq. (3)
from
$$\eta=0.069\biggl({A\over Z}\biggr)^{1/2}
\biggl({x\over\theta_c}\biggr)\biggl({
10^8\,{\rm T}\over B}\biggr)^{1/2}\biggl(
{P\over 1\,{\rm s}}\biggr),\eqno(10)$$
with $x\lapprox{\rm min}\{\theta_c, x_0\}$. Although we have a higher 
$\eta$ for the Arons \& Scharlemann model, the acceleration by the 
potential given by Eq. (3) is the more effective for $i\approx0$.

The evaluation of $x_0$ depends on the specific mechanism for
initiating a pair cascade and whether the pair plasma produced
through the cascade is dense enough to short
out part of the electric field. We consider three mechanisms
for starting a pair cascade: RICS, curvature radiation, and
inelastic scattering of ions by thermal photons from the
polar cap. The corresponding lengths are 
$x^{_{\rm RICS}}_0$, $x^{\rm curv}_0$ and $x^{\rm ion}_0$.
In general, we have $x^{_{\rm RICS}}_0<x^{\rm curv}_0$ (e.g.
Luo 1996). For moderately hot polar caps with  effective
temperature $\sim 10^6\,\rm K$, the energy loss due to
RICS is not important compared to acceleration but the photons
produced through RICS can start a pair cascade at the distance
less than $x^{_{\rm RICS}}_0<x^{\rm curv}_0$, $x^{\rm ion}_0$. Thus,
the effect of RICS is to reduce the gap acceleration efficiency.
For space-charge-limited flow, the composition of outflowing
charges at the poles with  $\mbox{\boldmath{$\Omega$}}\cdot{\bf B}>0$ 
and $\mbox{\boldmath{$\Omega$}}\cdot{\bf B}<0$ can be different. For 
$\mbox{\boldmath{$\Omega$}}\cdot{\bf B}>0$, the primary particles consist 
mainly of electrons, and for $\mbox{\boldmath{$\Omega$}}\cdot{\bf B}<0$, 
the main components are heavy ions or positrons. In the ion zone, 
the gap length is
constrained by the pair production by positrons through RICS.  The
possible source of positrons was discussed by Cheng \& Ruderman (1977).
For moderately hot polar caps, the gaps at both types of pole have a 
similar efficiency.

For sufficiently hot polar caps and a superstrong magnetic field
($\sim 10^9\,\rm T$), the energy loss of electrons or positrons
due to RICS can be important, and may prevent them starting a
pair cascade in the region close to the polar cap.  Since
the cross section is $\propto (Z^4/A^2)(m_e/m_p)^2\approx 10^{-4}$
for iron nuclei ($Z/A\approx1/2$), the energy loss due to RICS for
ions is negligibly small, and they can be continuously accelerated
until distance $x\lapprox{\rm min}\{x^{_{\rm RICS}}_0,x^{\rm ion}_0\}$.
Thus, the acceleration efficiency increases significantly.
As an example, for the potential (2) with $B=3\times10^{13}$
and $T=1.5\times10^6\,\rm K$, the gap
length is about $x^{_{\rm RICS}}_0R_0\approx 3\times10^3\,\rm cm$
and the maximum energy of electrons (or positrons) is $10^4m_ec^2$.
This gives the particle luminosity
$L_{\rm p}\approx 10^4m_ec^2\dot{N}_{_{\rm GJ}}$
where $\dot{N}_{_{\rm GJ}}$ is the injection rate of primary electrons
(or positrons). For hot polar caps with $T=10^7\,\rm K$, from Figures 1, 
we obtain $x^{_{\rm RICS}}_0R_0\approx 1.6\times10^4\,{\rm cm}$.
From Figure 4, pair production by ions interacting with thermal photons
is important only for the case where the whole star's surface has
$T=5\times10^6\,\rm K$ (the thick solid curve), and this occurs at the 
distance $x^{\rm ion}_0R_0\approx 10^5\,\rm cm$.  Thus, the gap length 
is controlled by RICS. From Figure 1, we have 
$\gamma(x^{_{\rm RICS}}_0)\approx 1.6\times10^5$ for positrons
(or electrons) and $\gamma(x^{_{\rm RICS}}_0)\approx(Zm_e/Am_p)\times10^6$
for ions; positrons (or electrons) and ions can be accelerated to the 
maximum energy $1.6\times10^5m_ec^2$ and $10^6 Zm_ec^2$, respectively. 
Hence, we have much a higher particle luminosity 
$L_{\rm p}\approx 10^6m_ec^2\dot{N}_{_{\rm GJ}}$
and most of $L_{\rm p}$ are carried by ions.

\section{Application to young pulsars}

From observations, several pulsars may have polar caps with
effective temperature $\ge10^6\, \rm K$ though the
exact value of the temperatures is rather uncertain
(for a review, see e.g., \"Ogelman 1991). For moderately hot polar caps
and typical B fields ($B\sim10^8\,\rm T$), the energy loss rate is
not rapid enough to compete with the acceleration, and RICS controls
the polar gap only through pair production. This in general reduces the
gap length, and hence reduces the acceleration efficiency by at
least one order magnitude, imposing a severe constraint on the energetics of
the polar gap.

For young pulsars with superstrong magnetic fields $\gapprox 10^9\ \rm T$, 
the gap acceleration efficiency can increase provided that the 
polar cap ($\mbox{\boldmath{$\Omega$}}\cdot{\bf B}<0$) is hot 
($T\ge5\times10^6\,\rm K$) and for electrons 
(or positrons), the energy loss due to RICS can dominate over the 
acceleration. One of CGRO pulsars, PSR 1509-58, has magnetic field
$\gapprox 1.5\times10^9\ \rm T$ (the magnetic field at the
pole can be twice this value). Since neutron stars are hot at birth
(after supernova explosion), their polar caps can be hot as result of
self-sustained polar cap heating at very young age (e.g. Luo 1996),
e.g within hundred years or so after birth. These high field pulsars,
assuming they were born with high magnetic fields, must have undergone a phase
when thermal emission had an important effect on pulsar electrodynamics.

We suggest that for high field pulsars there can be four phases
during the their lifetime as discussed below. (i) At very young age
(within a few years after supernova explosion) when the surface
or polar cap temperature is high, energy loss due to RICS is so
severe that a pair cascade due to electrons or positrons cannot occur
in the region close to the polar cap. In the ion zone 
($\mbox{\boldmath{$\Omega$}}\cdot{\bf B}<0$), accelerated
ions may carry most of the particle luminosity. For heavy ions,
photodisintegration may occur in the thermal photon field from
the polar cap. At this stage, since pulsars also have a very large spin-down
luminosity, they are a potential source of high energy hadrons
and electromagnetic radiation.
(ii) As pulsars cool down, the energy loss rate due to RICS becomes
less than acceleration rate. However, RICS may still be the dominant
mechanism for pair production to constrain the gap length (Luo 1996).
The high field pulsar, PSR 1509-58, may currently be at this stage
since the polar cap temperature inferred from the observation
is about $1.5\times10^6\,\rm K$ (e.g. Kawai et al. 1991).  
Its gap length is shorter than that controlled by pair production
by either curvature radiation or relativistic ions in the photon field.
Thus, the acceleration efficiency is less than
in phase (i). (iii) As the surface temperature or polar cap
temperature becomes much lower, pair production by RICS is not
important. The main mechanism for initiating pair cascades is
curvature radiation. As the period increases,
the gap length increases and hence the efficiency also increases.
However, at this stage, the pulsar spin-down luminosity is
lower than at their early young age. (iv) The period reaches
the critical period at which pair production due to curvature radiation
is no longer operative. Although the gap acceleration efficiency
is now high (no pairs to screen the electric field), the luminosity
is very low.

In summary, for very young pulsars with superstrong magnetic fields
($B\gapprox 10^9\ \rm T$) and hot polar caps ($\ge5\times10^6\,\rm K$),
the severe energy loss due to resonant inverse Compton scattering can
prevent electrons or positrons starting a pair cascade in the region
close to the polar cap but ions can be accelerated by the full potential
drop across the gap (provided that free emission of ions can occur).  
We then have the ions carrying most of the particle
luminosity.  Thus, very young pulsars with superstrong $B$ field can 
be a potential source of high-energy hadrons and gamma-rays. In contrast, 
for pulsars with moderately hot polar caps ($T\sim 10^6\,\rm K$), the 
effect of resonant inverse Compton scattering is to reduce the gap 
acceleration efficiency. Pair production by accelerated ions interacting
with thermal photons strongly depends on $P$ and $T$, and in general
we have $x^{_{\rm RICS}}_0<x^{\rm ion}_0$ and hence the polar gap
length is controlled by RICS.




\section*{Acknowledgements}

QL thanks Australian Research Council (ARC) for
financial support through a fellowship.
We thank Wlodek Bednarek for help discussion.


\section*{References}






\reference Arons, J. 1996, A\&AS, 120, 49.
\reference Arons, J. 1983, ApJ, 266, 215.
\reference Arons, J. \& Scharlemann, E. T. 1979, ApJ, 231, 854.
\reference Bednarek, W. \& Karakula, S. 1995, in
Proc. 24th Int. Cos. Ray Conf. (Rome), V.2, p. 279.
\reference Cheng, A. F. \& Ruderman, M. A. 1977, ApJ, 214, 598.
\reference Cheng, K. S., Ho, C. \& Ruderman, M. A. 1986, ApJ, 300, 500.
\reference Dermer, C. D. 1990, ApJ, 360, 197.
\reference Fawley, W. M., Arons, J. \& Scharlemann, E. T. 1977, ApJ, 217, 227.
\reference Greiveldinger, C. et al. 1996, ApJ, 465, L35.
\reference Herold, H. 1979, Phys. Rev., D19, 2868.
\reference Kawai, N. et al. 1991, ApJ, 383, L65.
\reference Luo, Q. 1996, ApJ, 468, 338.
\reference Michel, F. C. 1974, ApJ, 192, 713.
\reference \"Ogelman, H. 1991, in Thermal Emission of Pulsars, in Neutron 
           Stars: Theory and Observation, ed. J. Ventura \& D. Pines
           (Dordrecht: Kluwer), 87.
\reference Protheroe, R. J. 1984, Nature, 310, 296.
\reference Protheroe, R. J. 1997, in Towards the Millennium in 
           Astrophysics: Problems and Prospects, 
           eds. M.M. Shapiro and J.P.  Wefel 
           (World Scientific, Singapore), in press.
\reference Protheroe, R. J. \& Johnson, P. A. 1996, 
           Astroparticle Phys., 4, 253.
\reference Romani, R. W. 1987, ApJ, 313, 718.
\reference Ruderman, M. \& Sutherland, P. G. 1975, ApJ, 196, 51.
\reference Sturner, S. J. 1995, ApJ, 446, 292.
\reference Thompson, D. J. et al. 1994, ApJ, 436, 229.
\reference Ulmer, M. P. 1994, ApJSS, 90, 789.
\vfil

\end{document}